\documentclass[12pt]{article}
\usepackage[width=6.5in]{geometry}
\usepackage{bm, graphicx, array, verbatim, url, subcaption, booktabs, multirow, amsmath, amssymb, lineno, hyperref,xcolor,authblk}
\usepackage[inline]{enumitem}
\usepackage[font=footnotesize,skip=2pt]{caption}
\newcommand{\nip}{$\mathrm{Ni_{2}MnGa}$}
\newcommand{\nma}{$\mathrm{Ni_{2}MnAl}$}
\newcommand{\cma}{$\mathrm{Co_{2}MnAl}$}
\newcommand{\nix}{$\mathrm{Ni_{2-\mathit{x}}Q_\mathit{x}MnGa}$}
\newcommand{\nnmg}{$\mathrm{Ni_{2+\mathit{x}}Mn_{1-\mathit{x}}Ga}$}
\newcommand{\nixy}{$\mathrm{Ni_{2-\mathit{x}}Q_\mathit{x}MnGa_{1-\mathit{z}}R_z}$}
\newcommand{\nicx}{$\mathrm{Ni_{2-\mathit{x}}Co_\mathit{x}MnGa_{1-\mathit{z}}Mn_\mathit{z}}$}
\newcommand{\nmgm}{$\mathrm{Ni_2MnGa_{1-\mathit{z}}Mn_\mathit{z}}$}
\newcommand{\nmg}[1]{{{NiMnGa}}($#1$)}

\begin{document}
\title{\textit{Ab-initio} study of tuning the electronic and magnetic properties of \nip~Heusler alloy by Co and Mn compound doping}
\author{Karunakaran  M}
\author{Rudra Banerjee}
\affil{Department of Physics and Nanotechnology, SRM Institute of Science and Technology,
	Kattankulathur, Tamil Nadu,603203, India}
\maketitle
\begin{abstract}
	We report the effects of Mn and Co doping on the electronic properties, magnetic exchange interaction, and Curie temperature of
	intermetallic \nip~by Green's function based Korringa-Kohn-Rostoker method with coherent potential
	approximation(KKR-CPA). The effect of single and compound doping of Mn and Co at different crystallographic positions on these
	properties are evaluated by computing the electronic and magnetic structures of \nip.
	The study revealed the possibility of tuning magnetic exchange interaction($\mathcal{J}_{ij}$) and Curie temperature($T_C$)
	upon doping.
	Moreover, it is noted that doping can stabilize the Jahn-Teller distortion.
	It is also worth noting that $T_C$ responds in a different way with concentration as well as the site of the dopant.
	This study helps in understanding and realizing the cause for magnetic properties in \nip, and experimental peers can also
	use it for further research on doped \nip.
\end{abstract}
\section{\label{sec:intro}Introduction}
Energy is the necessary evil of the modern civilization. Reducing the production of greenhouse gas while keeping on with the
ever-increasing energy demand is a great challenge to mankind. Renewable energy technologies are one of the most trusted and tested
methods to limit greenhouse emissions and global warming. For caloric materials, which is one of the major parts
of renewable energy, the efficiency  depends on ordering temperature, e.g., Curie temperature($T_C$) for magnetocaloric (MC)
materials. Hence, to achieve the goal of 4/5$^{th}$ of the world's electricity by 2050 \cite{irena} from renewable sources, it is
extremely important to be able to tune the $T_C$ of the related materials \cite{Fitta_2018}.

Since the theoretical prediction of half-metallic properties in  $\mathrm{NiMnSb}$ by Groot \cite{de_Groot_1983} and experimental
observations by \cite{Ishida_1982,Fujii_1990,Jourdan_2014}, Heusler alloys(HA's) have drawn massive attention from the scientific
community \cite{Kelvin_2021, Sicong_2021, Wollmann_2017} ascribed to their wide usage in MC devices
\cite{Ahmad_2021}, magnetic shape memory alloys \cite{Yang_2019}, spintronics \cite{Bainsla_2015}, and giant magnetoresistance
devices \cite{Kundu_2017,Kihara_2021,Liu_2012}.  The full HA's, with generic symbol $X_2YZ$ stabilizes in the $L2_1$ structure
\cite{Mahat_2021} with a completely ordered phase. The $Y-Z$ disordered structure has the $B_2$ ground state \cite{Guan_2021}, while the
complete $X-Y-Z$ disordered system has the $A_2$ \cite{Vinicius_2021}.  Wyckoff positions of 4a(0, 0, 0),
4b($\frac{1}{2}, \frac{1}{2}, \frac{1}{2}$), and 8c($\frac{1}{4}, \frac{1}{4},\frac{1}{4}$) are occupied by {$Z$}, {$Y$}, and
	{$X$}, respectively \cite{Yang_2019, Heusler_1903, Heusler_1903_220}.
The ternary full HA, \nip, exhibits  both magnetic and structural phase transition \cite{Dilmieva_2017}. In this
context, Ni-Mn-Ga systems show
properties like shape memory effect and magnetic field-induced strain, which is advantageous in actuators and sensors
\cite{Yang_2019, Segui_2021}. They also show favorable MC (both conventional and inverse) properties, suitable to replace the century-old
Joule-Thomson cooling \cite{Kamantsev_2015, Franco_2012}.  Ni-Mn-Ga system exhibits magneto structural phase transition with a
huge change in isothermal  magnetic entropy($\Delta S_m \approx -18 $~J kg$^{-1}$ K$^{-1}$)
but a relatively narrow range of working temperature in full-width half maximum($T_{fwhm}$) around 290 K for an applied field
change of 5 T, typical to the materials with first-order phase transition(FOPT) \cite{Umetsu_2016}. However, the thermal
hysteresis of FOPT and the Curie temperature is predominant to magnetic refrigeration.


The electronic structure and magnetic properties have been reported for Ni-doped \cma~\cite{Okubo_2011} and Co-doped
\nma~\cite{Kanomata_2009} systems, with the dopant at the {$X$} position only. Off-stoichiometric \nix~ and \nixy,
where Q, and R are the dopant elements. For the past few decades, \nix/Al has been under study to
understand their martensite phase transition($T_M$) and Curie temperature($T_C$) \cite{Uijttewaal_2009,
	Sanjay_2017}.  In \nnmg, observed that the MC effect is the highest around $0.18\leqslant x\leqslant 0.27$ concentration as
magnetic and structural phase change occurs in this region \cite{Buchelnikov_2011}. Partially substituting Ni with Co is
generally known to enhance the ferromagnetic coupling and hence the $T_C$.  This increases the possibility that the quaternary
system undergoes a martensitic transition together with a meta-magnetic phase transition \cite{Kainuma_2006}.

Generally, the structural and magnetic properties are highly impacted by the off-stoichiometric combination of the main group
element($Z$) and transition metals($X$ and $Y$).  The effect of $Y$ replacing $Z$, mostly $\mathrm{Ni_2MnZ_{1-y}Mn}$, $Z$ = Ga,
In, etc., is well studied \cite{Kihara_2021,	Orlandi_2020, Sokolovskiy_2014}. Substituting the Mn element in the $Z$ position
can stabilize the cubic state of the system \cite{Kihara_2021}. The Mn$_{Z}$ atoms interact antiferromagnetically between the
surrounding Mn$_{Z}$ and normal Mn$_{Y}$ since the distance between Mn$_{Y}$-Mn$_{Z}$ is shorter than Mn$_{Y}$-Mn$_{Y}$
and Mn$_{Z}$-Mn$_{Z}$ \cite{Sokolovskiy_2012}. Here, Mn$_{Z}$ has been denoting the Mn atom in the Ga site($Z$), and Mn$_Y$ is
the Mn's normal position($Y$).  The doping of the Co element in the $X$ position can increase the ferromagnetic interaction
between the atoms and the Curie temperature of the material \cite{Kainuma_2006}. Additionally, even a small concentration of Co
atoms in the $X$ position significantly impacts the	Curie temperature.  Also, the cell volume effect and valence electron
concentration per atom ratio($e/a$) affect $T_C$ inversely, i.e.,  as the $e/a$ ratio increases, the $T_C$ decreases and
vice-versa \cite{Zhaoning_2018,Halder_2015}.

From the above discussion, investigation of the effect of doping both the $X$ and the $Z$ sites with $d$ elements, resulting to a disordered
\nicx, is interesting due to site preferences and their compound effect of electronic, magnetic, and thermodynamic properties. We
have investigated the effect of Co and Mn in the $X$ and $Z$ site and observed the effect in a restricted cubic phase.  The cubic
phase has minimum energy in the $L2_1$ structure. Increasing the Co concentration above 10\% retains
the cubic state as the ground state in a $B2$ structure \cite{Cao_2021}.  At around $x=0.25$, the cubic (austenite) phase collapses to
the tetragonal (martensitic) phase\cite{Kanomata_2009}.  We expected Mn doping in the Ga site and Co doping in the Ni site to enhance the
material's magnetic moment and Curie temperature.  The $x$ concentration varies from a value greater than 10\%, i.e., $x =
	0.12~\text{to}~0.24$.  Additionally, the $z$ concentration varies from $0~\text{to}~0.5$ since the $e/a$ ratio increases up to 8,
comparatively for the pure system \nip($e/a$ = 7.5). Therefore, the $z$ concentration is up to 0.5. This disorder system will be
referred to as \nmg{x, z} hereafter, where $x, z$ is the concentration of Co in the Ni site and Mn in the Ga site, respectively, for
brevity. For example, $\mathrm{Ni_{1.88}Co_{0.12}MnGa_{0.74}Mn_{0.26}}$ will be denoted as \nmg{0.12, 0.26}.

Notably, in this work, we have reported systematic studies of the variation of electronic structure, magnetic exchange
interactions, and tuning the Curie temperature($T_C$) of this disordered \nmg{x, z} system. We have especially looked into
patterns of the effect of variation of site occupancy upon a specific substitution for a wide range of concentrations. The
results are interpreted from the outcome of electronic structure calculations. This approach enables us to understand the
microscopic origin of the macroscopic property($T_C$).  The study of the substitution of a single $3d$ element on \nip~is, as
discussed above, very scattered. The investigation of double doping in \nip~is rare and far between. However, they offer
exciting phenomena, not only from the perspective of fundamental understanding but also for materials engineering with target
properties. For example, the tuning of $T_C$ is technologically relevant as it is one of the crucial factors that act as the
working temperature range of many devices, namely, MC, spintronics, etc. \cite{Ahmad_2021, Bainsla_2015}.

\section{\label{sec:methods}Methods}
The appropriate method for handling off-stoichiometric compositions is Green's function-based formalism with coherent potential
approximation(CPA),
as in the case of \nixy~\cite{BANERJEE_2010}.
We have performed \textit{ab-initio} calculations using multiple scattering Green's function formalism as implemented in
spin polarised relativistic Korringa-Kohn-Rostoker(SPRKKR)
code \cite{Ebert_2011,Ebert_2000, Mavropoulos_2006}.
The Perdew-Burke-Ernzerhof within generalized gradient approximation is used as the exchange-correlation
functional \cite{Perdew_1996}.
First Brillouin zone integrations were performed with 2500 grids of $k$-points,
and energy convergence criteria
were set as $10^{-5}$ Ry in the calculation in the range. We have implemented full potential spin-polarized scalar relativistic implementation
of SPRKKR with angular momentum cut-off $\ell_{max}=2$ as suitable for our system.

The lattice parameter with a minimum energy of \nmg{x,z} is calculated using the following procedure:
\begin{enumerate*}[label=(\roman*)]
	\item Obtain the lattice parameter from the materials project database \cite{Jain_2013};
	\item Calculated the self-consistent field of the system, with varying lattice parameter ranging from 94\% to 106\%, with identical calculation
	for each lattice parameters.
	\item Fit the lattice parameter vs energy plot obtained in the last step using a 4$^{th}$ order polynomial.
\end{enumerate*}
The parameter corresponding to the minima of the curve is the optimized lattice parameter. For \nmg{x, z}, we have taken the optimized
lattice parameter of previous calculations that has a minimum $x, z$ change as the starting point and followed the steps above. We
have shown the optimization curve of \nmg{0,0} in Figure (\ref{fig:lopt}). Other minimization energy curves are not
shown here for brevity.
In the present study, doping concentrations are varied by changing the Wyckoff site occupancy of the corresponding element in
the input structure. The resultant structure is used for further calculations.

The magnetic exchange energy($\mathcal{J}^{\nu\mu}_{ij}$) was calculated to understand the properties of magnetic interactions.
The Heisenberg model is defined as, \begin{equation} H=\sum_{\nu,\mu}\sum_{i,j}
	\mathcal{J}^{\nu\mu}_{ij}\mathbf{e}_i^\nu\cdot\mathbf{e}_j^\mu \end{equation} where $\nu,\mu$ represent atoms in different
sublattices, $i, j$ is a different lattice point, $\mathbf{e}^\nu_i$ is the magnetic orientation of $i^{th}$ atom at $\nu$
sublattices. The $\mathcal{J}_{i,j}^{\nu\mu}$ is calculated by the energy difference due to an infinitesimal change of magnetic
direction, as formulated by Lichtenstein \cite{Liechtenstein_1987}.


Finally, the $T_C$ is estimated using mean field theory, yielding
\begin{equation}
	k_BT_C= \frac{3}{2} \mathcal{J}^{\nu\mu}_{l}
\end{equation}
where $\mathcal{J}_l$ is the largest eigenvalue of the determinant, as described in \cite{Sokolovskiy_2012}. It must be
remembered that mean field calculations generally overestimate the $T_C$.

\section{\label{sec:results}Results}
\begin{figure}[ht!]
	\centering
	\begin{subfigure}[t]{.23\columnwidth}
		\includegraphics[width=0.9\linewidth]{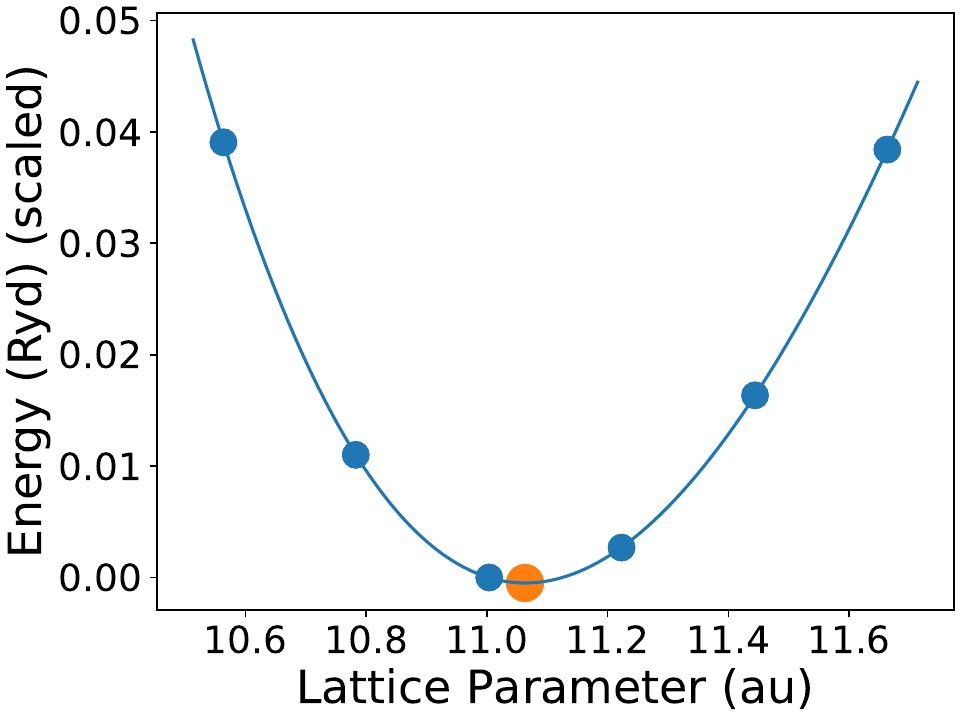}
		\caption{\label{fig:lopt}Energy optimization of \nip. }
	\end{subfigure}
	\begin{subfigure}[t]{.23\columnwidth}
		\includegraphics[width=0.9\linewidth]{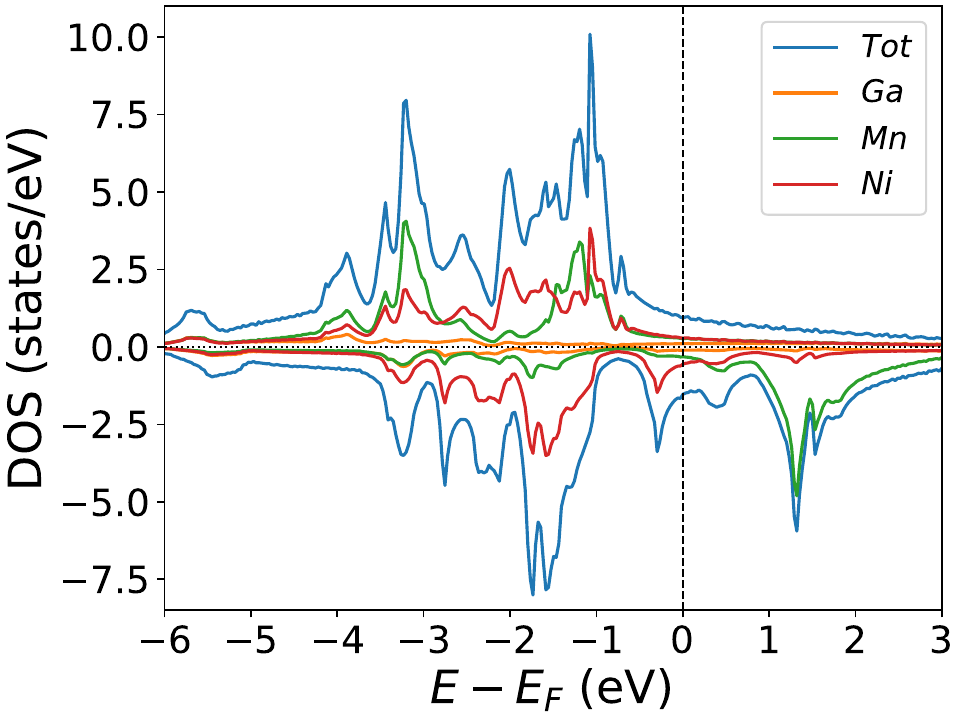}
		\caption{\label{fig:DOS_0_0} DOS of \nip.}
	\end{subfigure}%
	\begin{subfigure}[t]{.23\columnwidth}
		\includegraphics[width=0.9\linewidth]{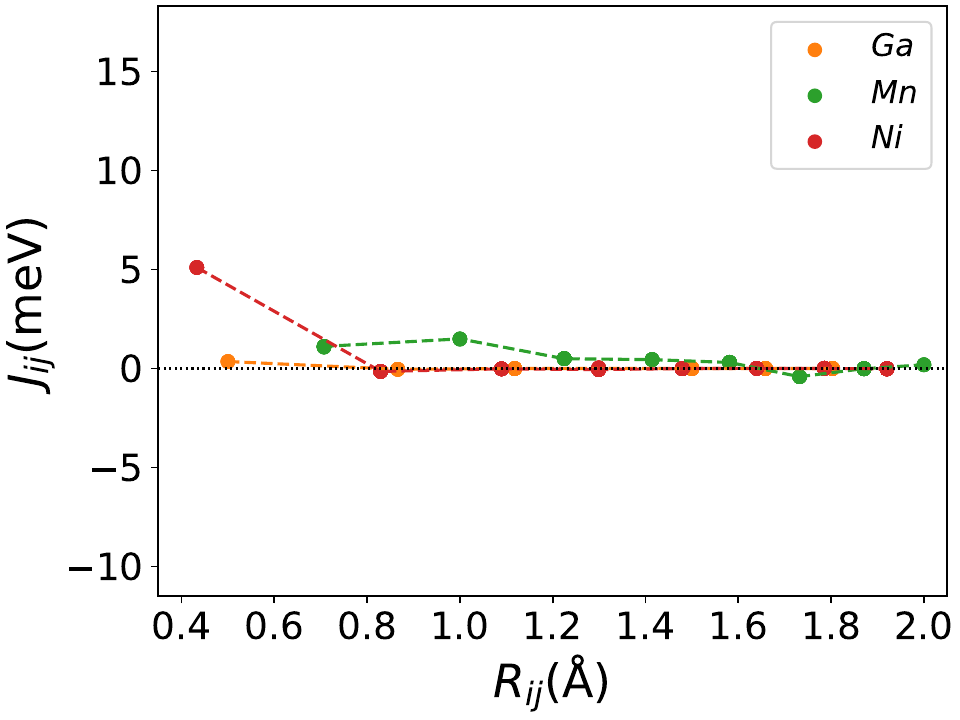}
		\caption{\label{fig:Jij_0_0}$\mathcal{J}_{ij}$ of \nip.}
	\end{subfigure}
	\begin{subfigure}[t]{.23\columnwidth}
		\includegraphics[width=1.17\linewidth]{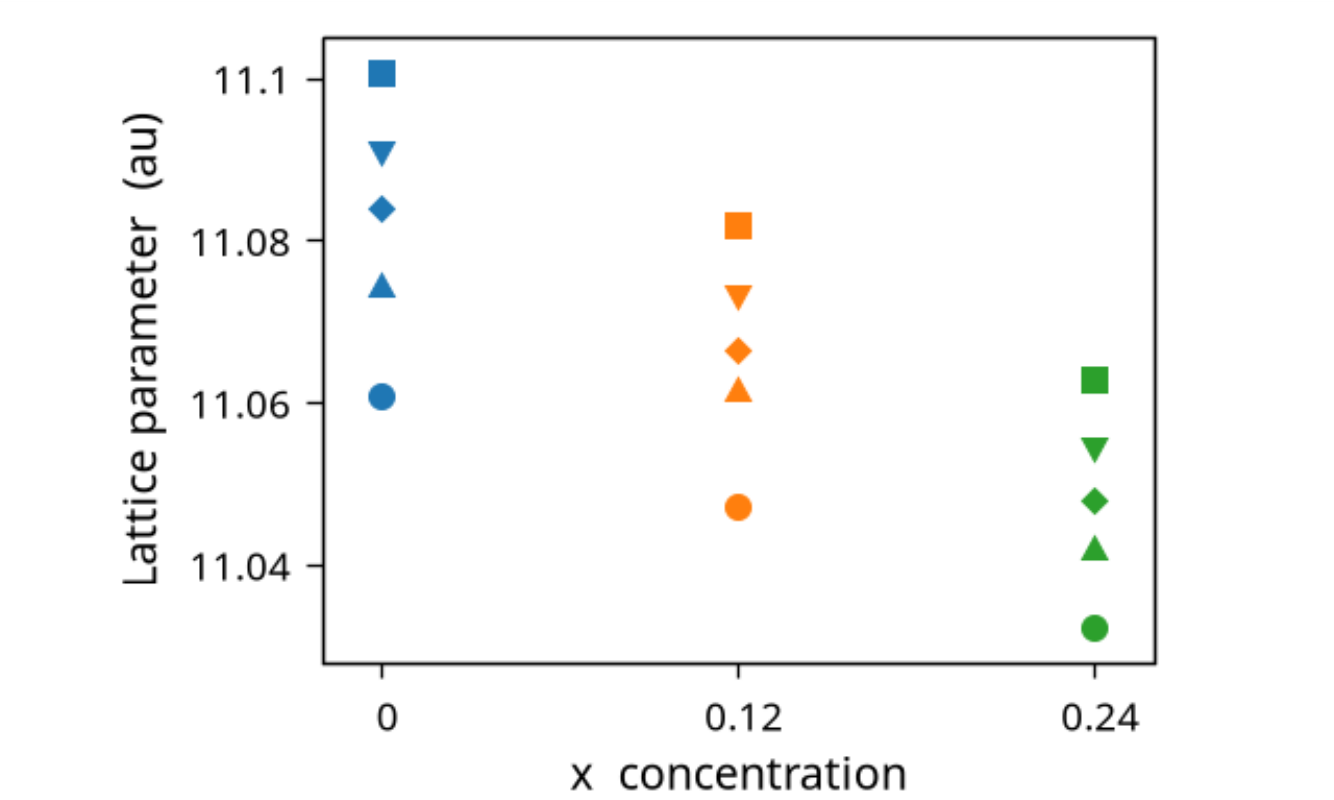}
		\caption{\label{fig:lvar}Variation of lattice parameter of \nmg{x,z}}
	\end{subfigure}
	\caption{Structure, electronic and magnetic properties of \nip. Figure (\ref{fig:lopt}) shows the energy/volume optimization of
		\nip. The blue dots are the actual calculation points; the orange dot is the lattice parameter for minimal energy. Figure
		(\ref{fig:DOS_0_0}) and Figure (\ref{fig:Jij_0_0}) shows the total and atoms projected DOS and magnetic pair interactions
		respectively. Figure (\ref{fig:lvar}) shows the variation of lattice parameters as a function of doping concentration at the
		$X$ and $Z$ sites. Here, the symbol's colors blue, orange, and green represent the values of $x = 0,
			0.12,~\text{and}~0.24$. The filled symbols of circle, upward triangle, diamond, downward triangle, and square are
		indicating the $z = 0, 0.15, 0.26, 0.35,~\text{and}~0.50$, respectively.}
	\label{fig:lcal}
\end{figure}

We have calculated the electronic and magnetic properties of \nmg{x, z} for ($x=0.0, 0.12\text{~and~}0.24$) and ($z=0.0, 0.15,
	0.26, 0.35\text{ and~}0.5$).  The optimized lattice parameter of each sample was calculated using the method described in the
sec.  (\ref{sec:methods}) and tabulated in Tables (\ref{tab:nmg_x-0}, \ref{tab:nmg_0-y}, \ref{tab:nmg_x-z}).
Figure (\ref{fig:lvar}) shows the variation of lattice parameters with $x$ and $z$, which is mostly linear.

Figures (\ref{fig:lopt}-\ref{fig:Jij_0_0}) shows the calculations of the pure \nip. Figure (\ref{fig:lopt}) shows the
lattice parameter optimization as described in section (\ref{sec:methods}). Our calculated density of states(DOS)
(Figure (\ref{fig:DOS_0_0})) and magnetic exchange interactions($\mathcal{J}_{ij}$) (Figure (\ref{fig:Jij_0_0})) with the Mn atom at
the center for \nip~matches the previous findings \cite{Sahariah_2012}. The $\mathcal{J}_{ij}$ is highest for Ni-Mn interactions,
with the value $\approx$ 5 meV. All the interactions are predominantly ferromagnetic in this case.  The complete table of
optimized lattice parameters, the total and individual magnetic moment per atom, and Curie temperature of the pure system is
tabulated in Table (\ref{tab:nmg_x-0}).

The change of lattice parameter with $x,~z$ in the cubic domain is shown in Figure (\ref{fig:lvar}).
This trend shows that doping Co at the $X$ site decreases the lattice parameter but doping at the $Z$ site increases the lattice
parameter.

\subsection{\nmg{x,0} systems}\label{sec:x0_system}
\begin{figure}[ht!]
	\centering
	\begin{subfigure}{.48\columnwidth}
		\begin{subfigure}{.48\columnwidth}
			\renewcommand\thesubfigure{\alph{subfigure}.1}
			\includegraphics[width=1\linewidth]{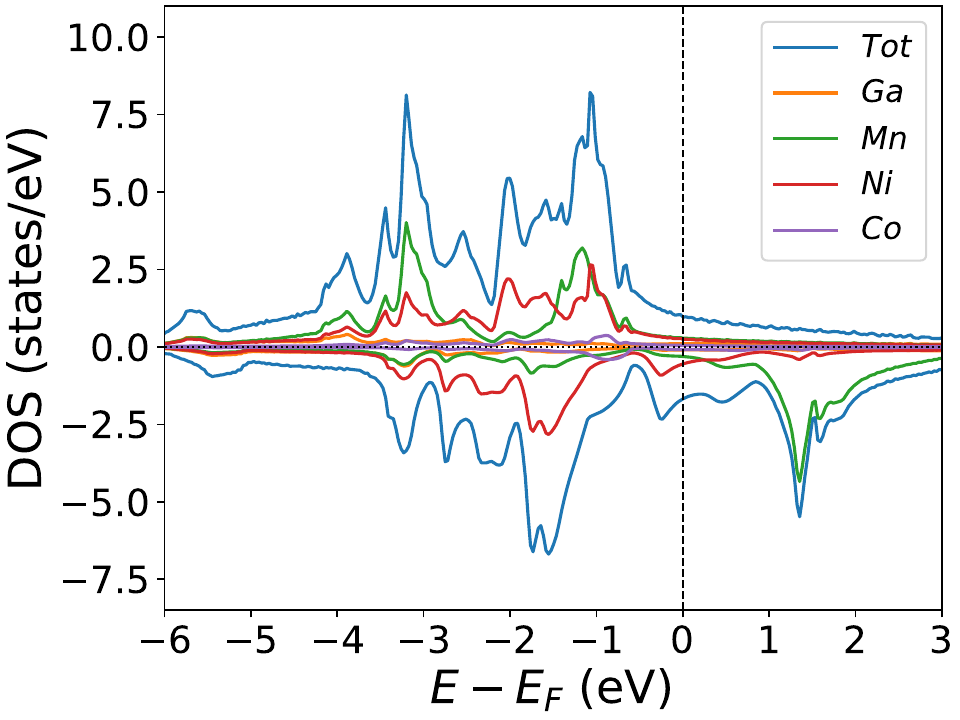}
			\caption{\label{fig:DOS_12_0}DOS of \nmg{0.12,0}.}
		\end{subfigure}%
		\begin{subfigure}{.48\columnwidth}
			\addtocounter{subfigure}{-1}
			\renewcommand\thesubfigure{\alph{subfigure}.2}
			\includegraphics[width=1\linewidth]{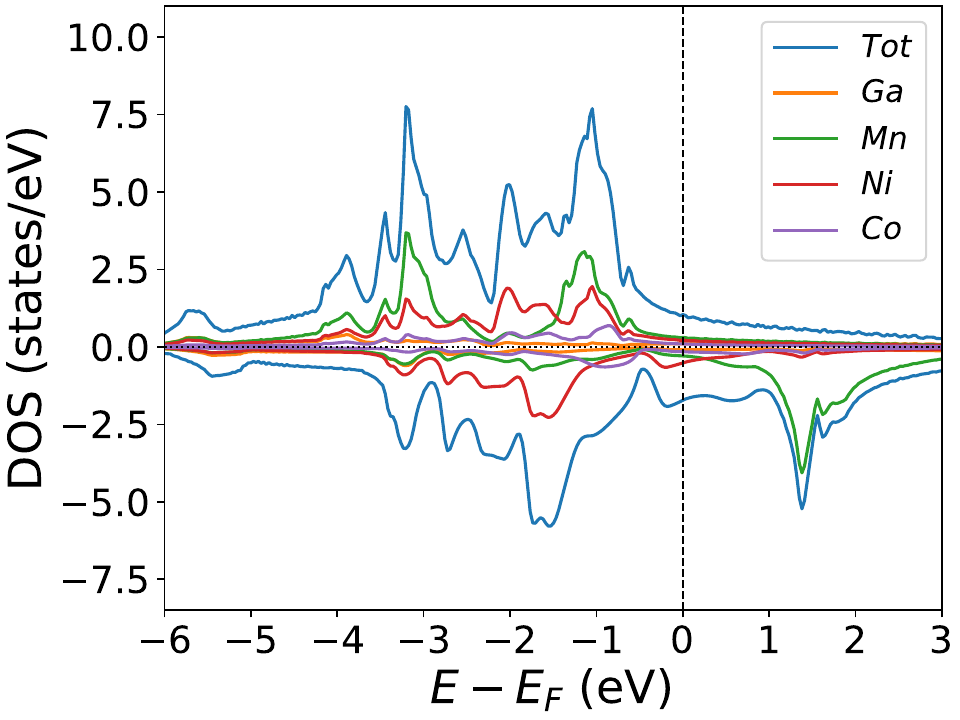}
			\caption{\label{fig:DOS_24_0}DOS of \nmg{0.24,0}.}
		\end{subfigure}
		\label{fig:nmg_x0-dos}
	\end{subfigure}
	\begin{subfigure}{.48\columnwidth}
		\begin{subfigure}{.48\columnwidth}
			\renewcommand\thesubfigure{\alph{subfigure}.1}
			\includegraphics[width=1\linewidth]{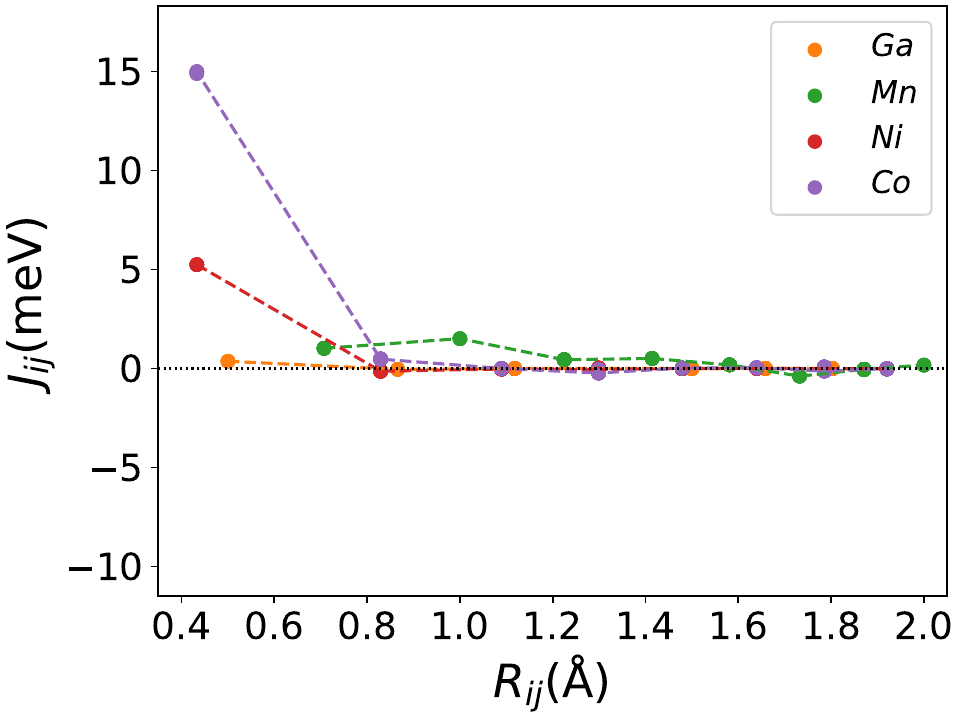}
			\caption{\label{fig:Jij_12_0}$\mathcal{J}_{ij}$ of \nmg{0.12,0}.}
		\end{subfigure}
		\begin{subfigure}{.48\columnwidth}
			\addtocounter{subfigure}{-1}
			\renewcommand\thesubfigure{\alph{subfigure}.2}
			\includegraphics[width=1\linewidth]{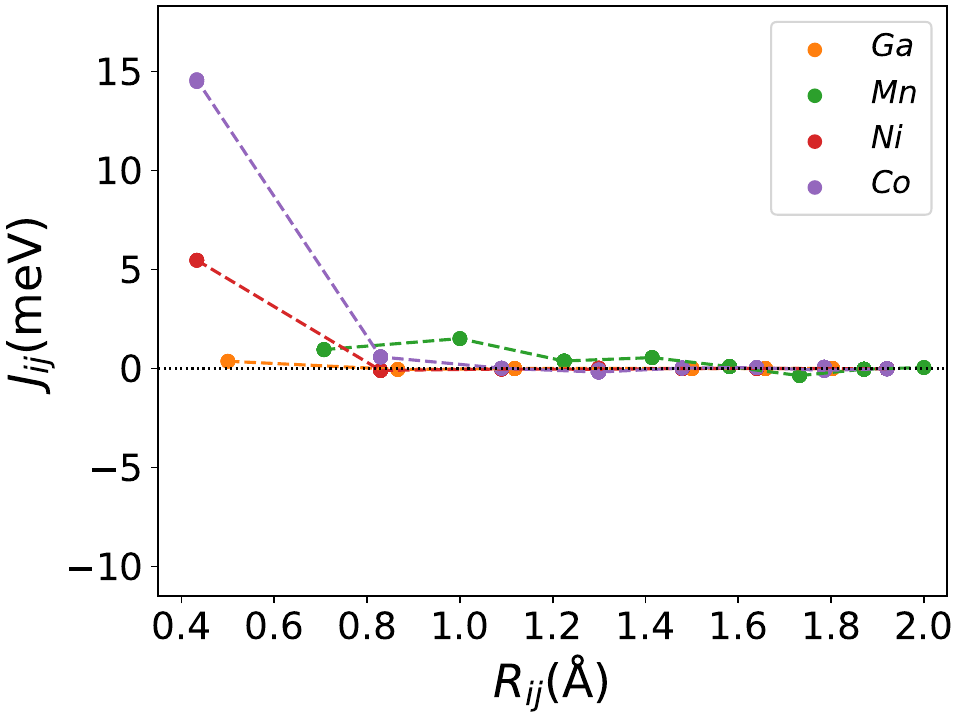}
			\caption{\label{fig:Jij_24_0}$\mathcal{J}_{ij}$ of \nmg{0.24,0}.}
		\end{subfigure}
		\label{fig:nmg_x0-jij}
	\end{subfigure}
	\caption{The (\subref{fig:nmg_x0-dos})DOS and (\subref{fig:nmg_x0-jij})exchange interaction of \nmg{x,0} with
		$x=0.12$ in Figure (\subref{fig:DOS_12_0},\subref{fig:Jij_12_0}) and $x=0.24$ in Figure
		(\subref{fig:DOS_24_0},\subref{fig:Jij_24_0}). Compared to DOS of pure \nmg{0,0},  (Figure (\ref{fig:DOS_0_0})), the DOS shows
		smeared peak just below the Fermi level and shrank pseudogap in minority spin channel with increase in $x$.
		Figure (\subref{fig:nmg_x0-jij}) shows the magnetic exchange interaction with the Mn$_{Y}$ atom at  the center. The magnetic
		interaction remains almost the same between $x=0.12$ and $x=0.24$ with the largest interaction
		between the Co and Mn.}
	\label{fig:nmg_x-0}
\end{figure}

The \nmg{x, 0} system, doping in Ni site only, though heavily studied within the austenite phase, we have included them for
completeness. The doping of Co in the $X$ position will decrease the lattice parameter of the material shown in
Figure (\ref{fig:lvar}) and Table (\ref{tab:nmg_x-0}).
The electronic and magnetic structures of \nmg{x,0} are shown in Figure (\ref{fig:nmg_x-0}) as representative of the series.
Generally, the $X$ and $Y$ site atoms contribute more to the net magnetic moment of the system.  Here the Mn atom
loses the moment due to the presence of an effective $X$ site(Ni with Co), and the Ni atom holds the magnetic moment  up to
the other elements present with the Ni atom. While replacing the Ni with Co in the $X$ site, the moment of Ni increases
up to 14\% for $z = 0.12~\text{and}~0.24$, as tabulated in Table (\ref{tab:nmg_x-0}).
The Co atom has a much higher magnetic moment($\approx 1~\mu_B$) than Ni($\approx 0.3~\mu_B$), and the $\mathcal{J}_{ij}$
is also much higher than \nip.
Increasing the Co concentration causes the magnetic moments of Co and Mn to start decreasing, and the magnetic
moment of Ni starts increasing, thereby enhancing the total magnetic moment of the system.
Of course, in the current case, Mn$_{Z}$ =0.

The magnetic properties depend substantially on the exchange interactions between all pairs of chemical elements.
In Figures (\ref{fig:Jij_0_0}), (\ref{fig:Jij_12_0}), (\ref{fig:Jij_24_0}), the exchange interaction is shown for each atom in
the configuration of $x~=~0,~0.12,~\text{and}~0.24$, and $z~=~0$.
From these Figures, the interactions between all pairs(Ni-Mn, Co-Mn, Mn-Mn, and Ga-Mn) are
ferromagnetic, and especially the Ni-Mn and Co-Mn are more ferromagnetic.
The interaction between Ni/Co
and Mn atoms in different sublattices is consistently ferromagnetic \cite{Ghosh_2010}.
In this way, the replacement of Ni with Co in the $X$ site enhanced the exchange interaction energy in Co-Mn pair, then the
ferromagnetic coupling also increased. These enhanced ferromagnetic coupling caused an increase in the Curie temperature of the
material. The pair of exchange coupling of Ni-Mn, Co-Mn, and Mn-Mn are more in the austenite phase than in the martensite phase. This
behavior  results in the decrease of the martensite temperature with increasing the Co concentration \cite{Sokolovskiy_2014}.

When increasing the Co concentration at the Ni site(with $z=0$) leads to a decrease in the $e/a$ ratio and the
lattice parameter(cell volume). This could be causing the $T_C$ to increase, as tabulated in the Table (\ref{tab:nmg_x-0}).

\begin{table*}[ht!]
	\centering
	\begin{tabular}{ccccccccccc}\toprule
		\multicolumn{2}{c}{Concentrations} & {Lattice} & \multicolumn{6}{c}{Magnetic Moment($\mu_B$)} & {$T_C$}  & \multirow{2}{*}{$e/a$}
		\\\cmidrule{4-9}
		$x$                                & $z$       & Parameter (au)                               & Ni       & Co
		                                   & Mn$_{Y}$  & Ga                                           & Mn$_{Z}$ & Total                  & (K)  &
		\\\toprule
		0                                  & 0         & 11.061                                       & 0.29     & -                      & 3.56
		                                   & -0.08     & -                                            & 4.08     & 393                    & 7.50   \\\midrule
		0.12                               & 0         & 11.047                                       & 0.31     & 1.07                   & 3.50
		                                   & -0.08     & -                                            & 4.21     & 1138                   & 7.44   \\
		0.24                               & 0         & 11.032                                       & 0.33     & 0.95                   & 3.45
		                                   & -0.09     & -                                            & 4.34     & 1148                   & 7.38
		\\\bottomrule
	\end{tabular}
	\caption{{\label{tab:nmg_x-0}}Structural, magnetic, and thermodynamic properties of \nmg{0, 0} and \nmg{x, 0} system.}
\end{table*}

\subsection{\label{sec:0,y_systems}\nmg{0, z} systems}
The opposite system of the one discussed in the previous section (\S\ref{sec:x0_system}) is \nmg{0, z} system, where the $X$ site is
fully ordered, but Ga is replaced by Mn, yielding \nmgm~system. The doping of Mn in the $Z$ site, will increase the lattice parameter
of the material(as shown in Figure (\ref{fig:lvar})) and induce the magnetic atom's moments as tabulated in Table
(\ref{tab:nmg_0-y}). Notably, the moment of Ni increases up to 75\% compared to the pure system, as discussed in the
earlier section	(\S\ref{sec:results}). The Mn$_{Y}$ magnetic moment increases linearly with the constant magnetic moment of
Mn$_{Z}$.

\begin{figure}[ht]
	\centering
	\begin{subfigure}{\columnwidth}
		\begin{subfigure}{.24\columnwidth}
			\renewcommand\thesubfigure{\alph{subfigure}.1}
			\includegraphics[width=1\linewidth]{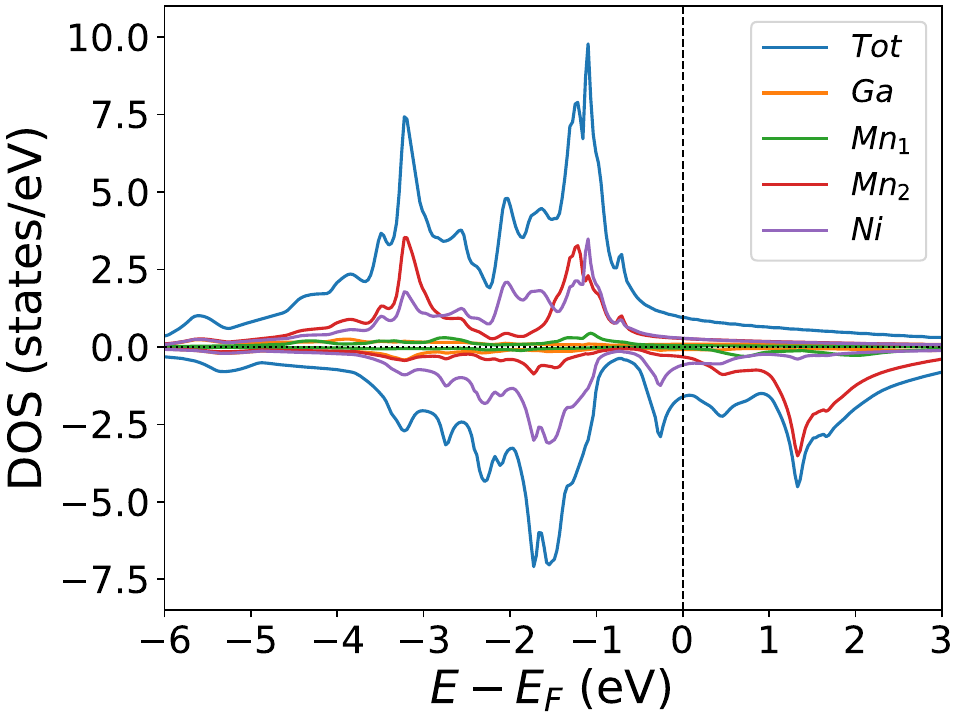}
			\caption{\label{fig:DOS_0_15}DOS of \nmg{0,0.15}.}
		\end{subfigure}%
		\begin{subfigure}{.24\columnwidth}
			\addtocounter{subfigure}{-1}
			\renewcommand\thesubfigure{\alph{subfigure}.2}
			\includegraphics[width=1\linewidth]{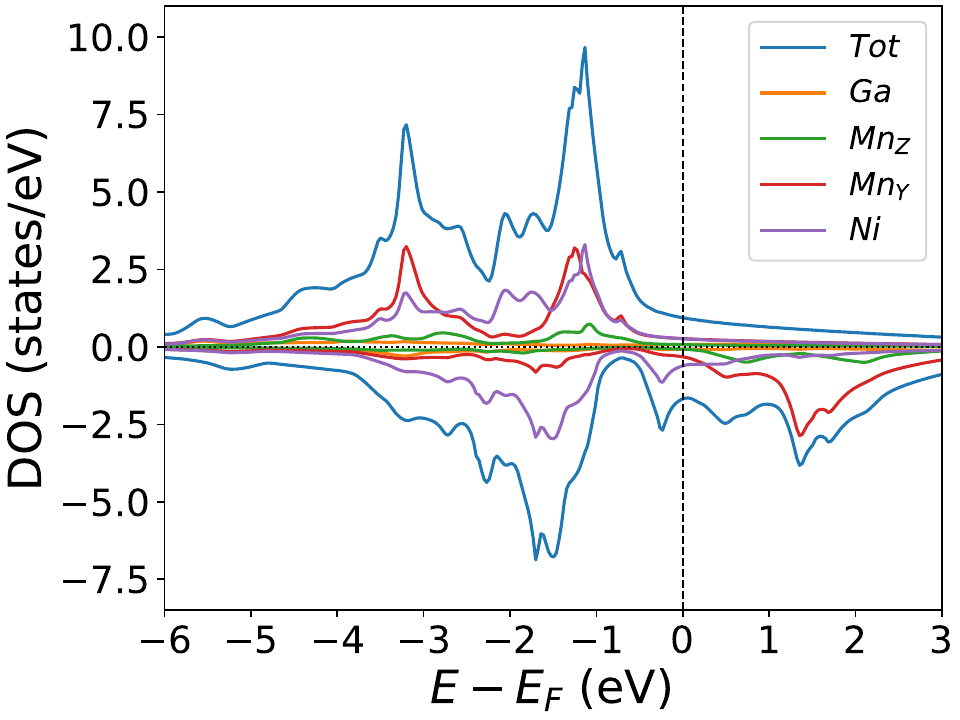}
			\caption{\label{fig:DOS_0_26}DOS of \nmg{0,0.26}.}
		\end{subfigure}
		\begin{subfigure}{.24\columnwidth}
			\addtocounter{subfigure}{-1}
			\renewcommand\thesubfigure{\alph{subfigure}.3}
			\includegraphics[width=1\linewidth]{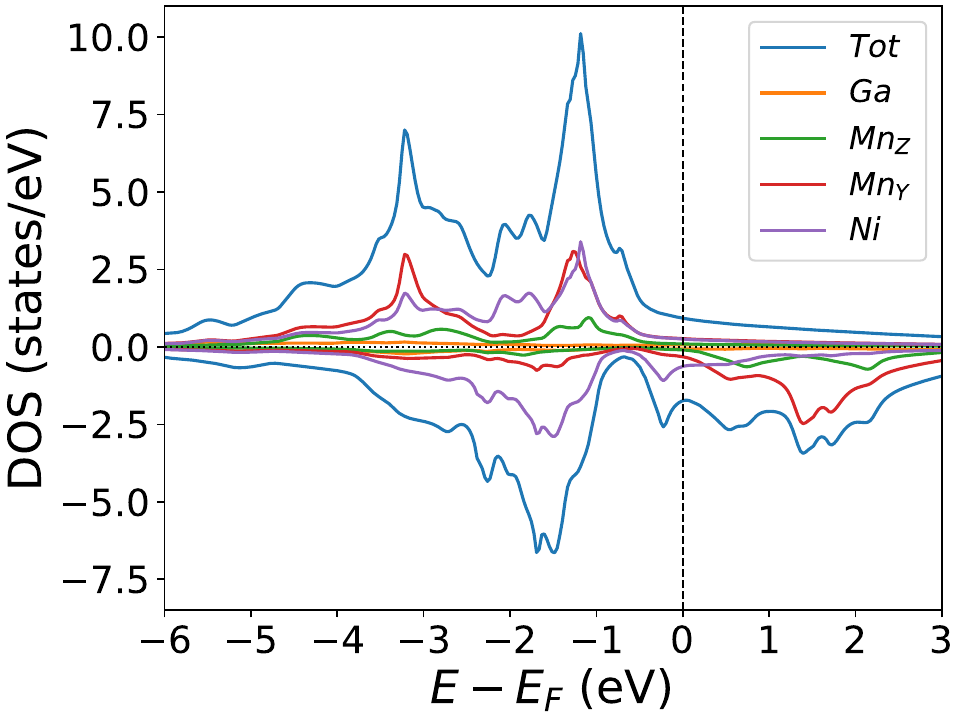}
			\caption{\label{fig:DOS_0_35}DOS of \nmg{0,0.35}.}
		\end{subfigure}
		\begin{subfigure}{.24\columnwidth}
			\addtocounter{subfigure}{-1}
			\renewcommand\thesubfigure{\alph{subfigure}.4}
			\includegraphics[width=1\linewidth]{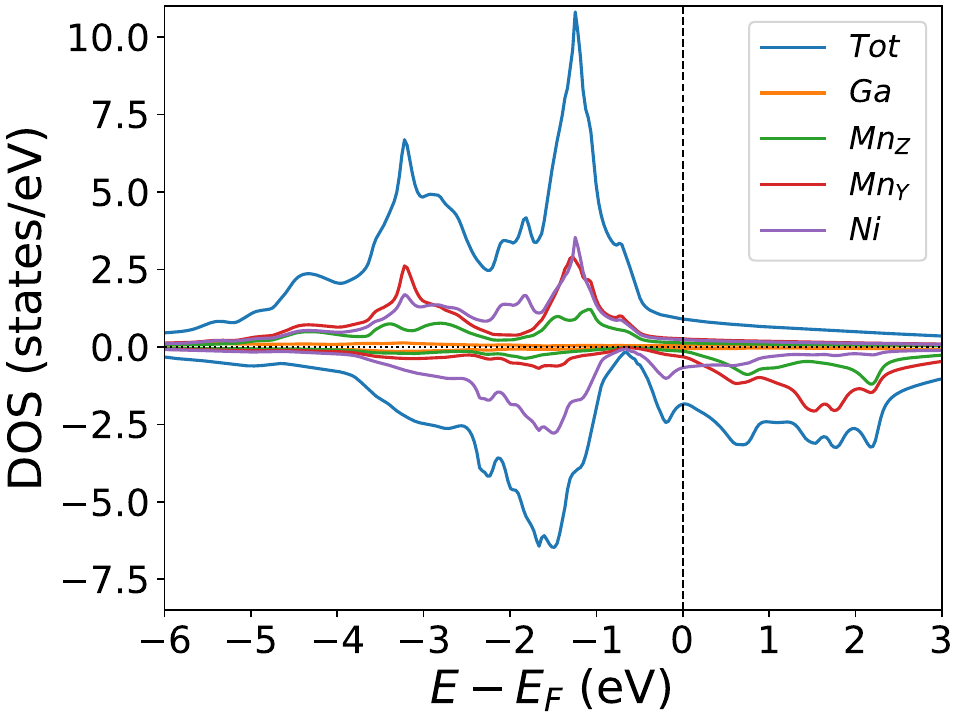}
			\caption{\label{fig:DOS_0_50}DOS of \nmg{0,0.50}.}
		\end{subfigure}
		\label{fig:nmg_0z-dos}
	\end{subfigure}
	\begin{subfigure}{\columnwidth}
		\begin{subfigure}{.24\columnwidth}
			\renewcommand\thesubfigure{\alph{subfigure}.1}
			\includegraphics[width=1\linewidth]{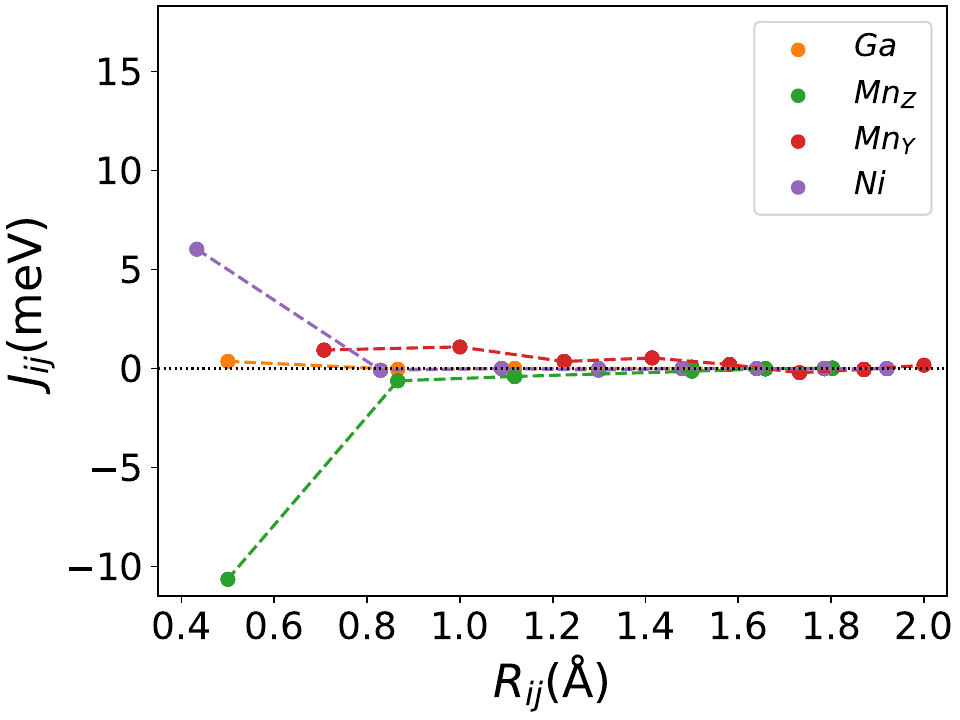}
			\caption{\label{fig:Jij_0_15}$\mathcal{J}_{ij}$ of \nmg{0,0.15}.}
		\end{subfigure}
		\begin{subfigure}{.24\columnwidth}
			\addtocounter{subfigure}{-1}
			\renewcommand\thesubfigure{\alph{subfigure}.2}
			\includegraphics[width=1\linewidth]{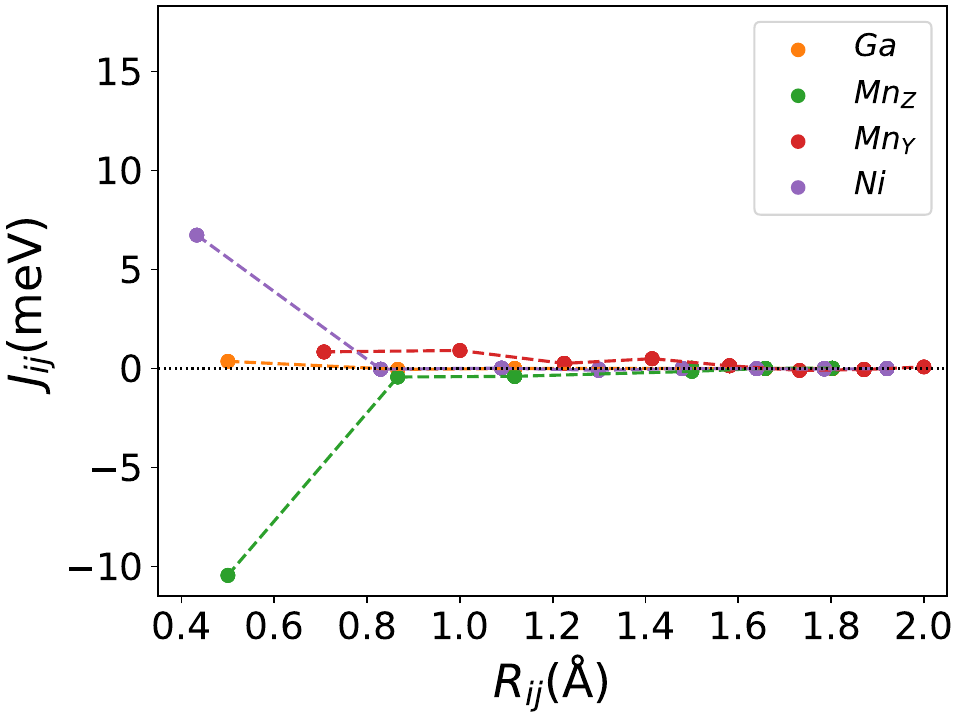}
			\caption{\label{fig:Jij_0_26}$\mathcal{J}_{ij}$ of \nmg{0,0.26}.}
		\end{subfigure}
		\begin{subfigure}{.24\columnwidth}
			\addtocounter{subfigure}{-1}
			\renewcommand\thesubfigure{\alph{subfigure}.3}
			\includegraphics[width=1\linewidth]{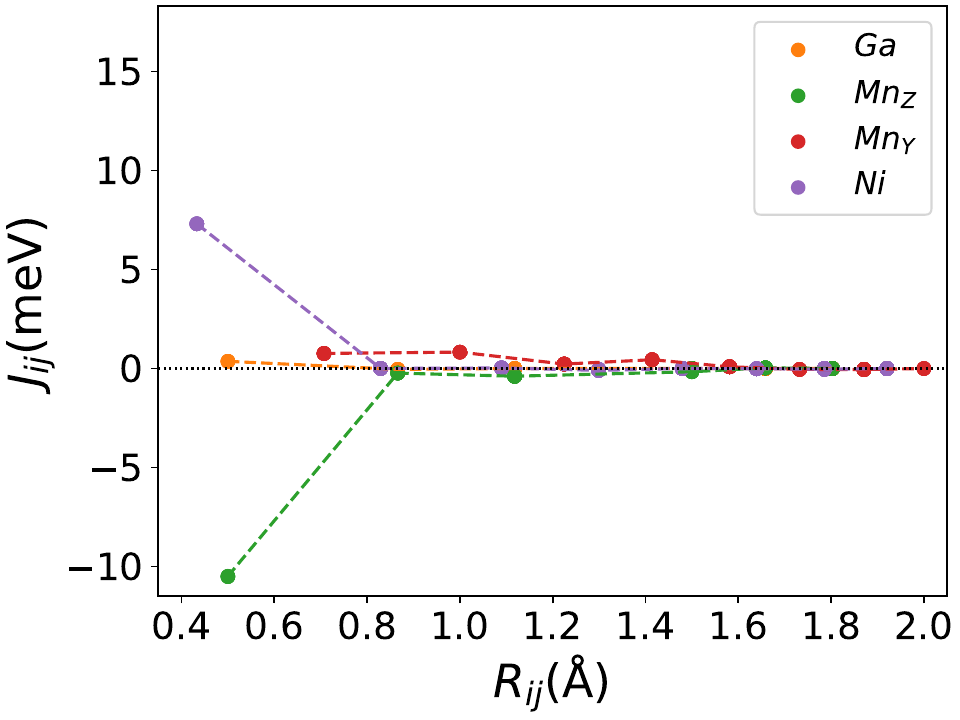}
			\caption{\label{fig:Jij_0_35}$\mathcal{J}_{ij}$ of \nmg{0,0.35}.}
		\end{subfigure}
		\begin{subfigure}{.24\columnwidth}
			\addtocounter{subfigure}{-1}
			\renewcommand\thesubfigure{\alph{subfigure}.4}
			\includegraphics[width=1\linewidth]{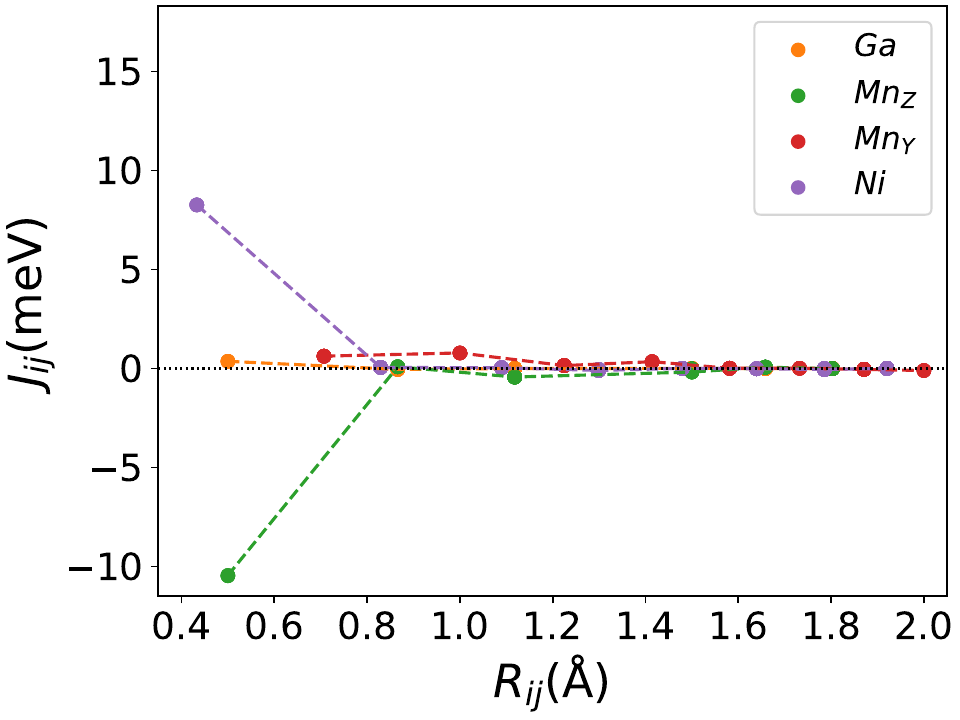}
			\caption{\label{fig:Jij_0_50}$\mathcal{J}_{ij}$ of \nmg{0,0.50}.}
		\end{subfigure}
		\label{fig:nmg_0z-jij}
	\end{subfigure}
	\caption{The (\subref{fig:nmg_0z-dos})DOS and (\subref{fig:nmg_0z-jij})exchange interaction of \nmg{0, z} ($z$ = 0.15, 0.26, 0.35
		and 0.50).}
	\label{fig:ydope}
\end{figure}

The electronic structure does not change much upon doping(Figure (\ref{fig:DOS_0_26})), even though the absence of Co in the Ni
site increases the magnetic interactions of Ni-Mn by half(Figure (\ref{fig:Jij_0_50})) with respect to in the case of
\nmg{x,0}(Figure (\ref{fig:nmg_x-0}\subref{fig:nmg_x0-jij})).
The most dominant interaction is Mn$_{Y}$-Mn$_{Z}$ interaction, which is antiferromagnetically coupled as shown in Figure
(\ref{fig:ydope}\subref{fig:nmg_0z-jij}) for each concentration.

Increasing the doping concentration of Mn in the Ga site(with $x=0$), accordingly the $e/a$ ratio starts to
increase since Mn has more valence electrons compared to Ga. Additionally, the lattice parameter starts increasing and
which leads to an increase in cell volume, while an increase in the Mn in the Ga site. This could be causing the $T_C$ to
start decreasing after an initial increase from the pure system. The values of the $e/a$ ratio, lattice parameter, and
$T_C$ as tabulated in the Table (\ref{tab:nmg_0-y}).

\begin{table*}[ht!]
	\centering
	\begin{tabular}{ccccccccccc}\toprule
		\multicolumn{2}{c}{Concentrations} & {Lattice} & \multicolumn{6}{c}{Magnetic Moment($\mu_B$)} & {$T_C$}  &
		\multirow{2}{*}{$e/a$}
		\\\cmidrule{4-9}
		$x$                                & $z$       & Parameter (au)                               & Ni       & Co
		                                   & Mn$_{Y}$  & Ga                                           & Mn$_{Z}$ & Total & (K)
		\\\toprule
		\multirow{3}{*}{0}                 & 0.15      & 11.074                                       & 0.35     & -     & 3.57 &
		-0.08                              & 3.65      & 4.76                                         & 741      & 7.65           \\
		                                   & 0.26      & 11.084                                       & 0.40     & -     & 3.58 &
		-0.08                              & 3.65      & 5.27                                         & 691      & 7.76           \\
		                                   & 0.35      & 11.091                                       & 0.44     & -     & 3.59 &
		-0.09                              & 3.65      & 5.69                                         & 664      & 7.85           \\
		                                   & 0.50      & 11.101                                       & 0.51     & -     & 3.60 &
		-0.10                              & 3.65      & 6.39                                         & 623      & 8.00           \\\bottomrule
	\end{tabular}
	\caption{\label{tab:nmg_0-y}Structural, magnetic and thermodynamic properties of \nmg{0,z} system.}
\end{table*}

\subsection{\label{sec:com_dis}Complete disorder: \nmg{x, z} systems}
\begin{figure}[ht]
	\begin{subfigure}{.24\columnwidth}
		\centering
		\includegraphics[width=1\linewidth]{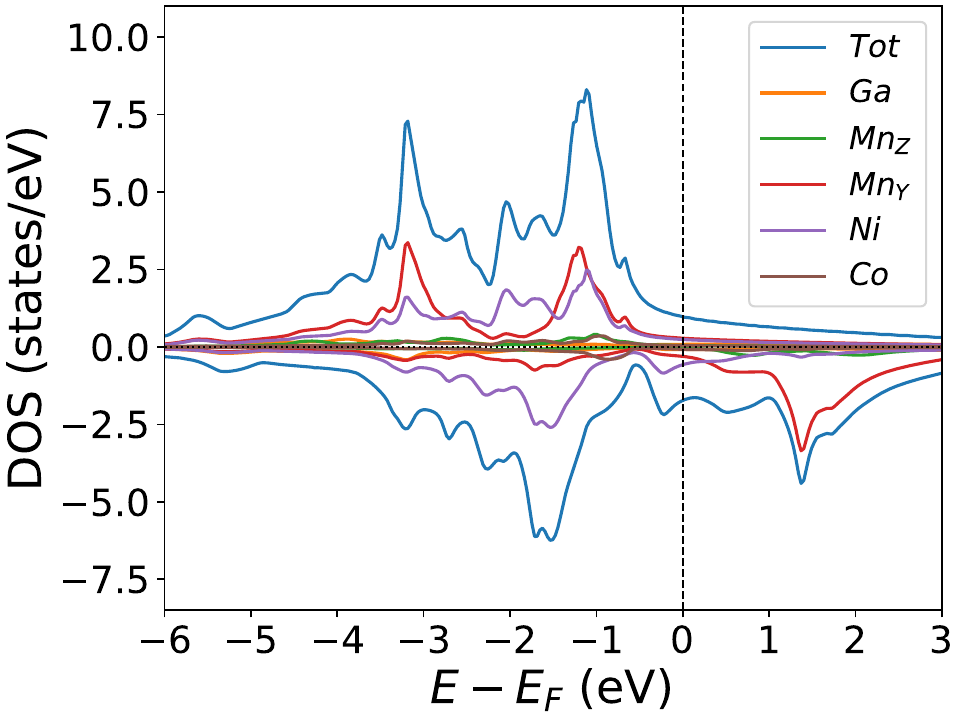}
		\caption{\label{fig:DOS_12_15}DOS of \nmg{0.12,0.15}.}
	\end{subfigure}%
	\begin{subfigure}{.24\columnwidth}
		\centering
		\includegraphics[width=1\linewidth]{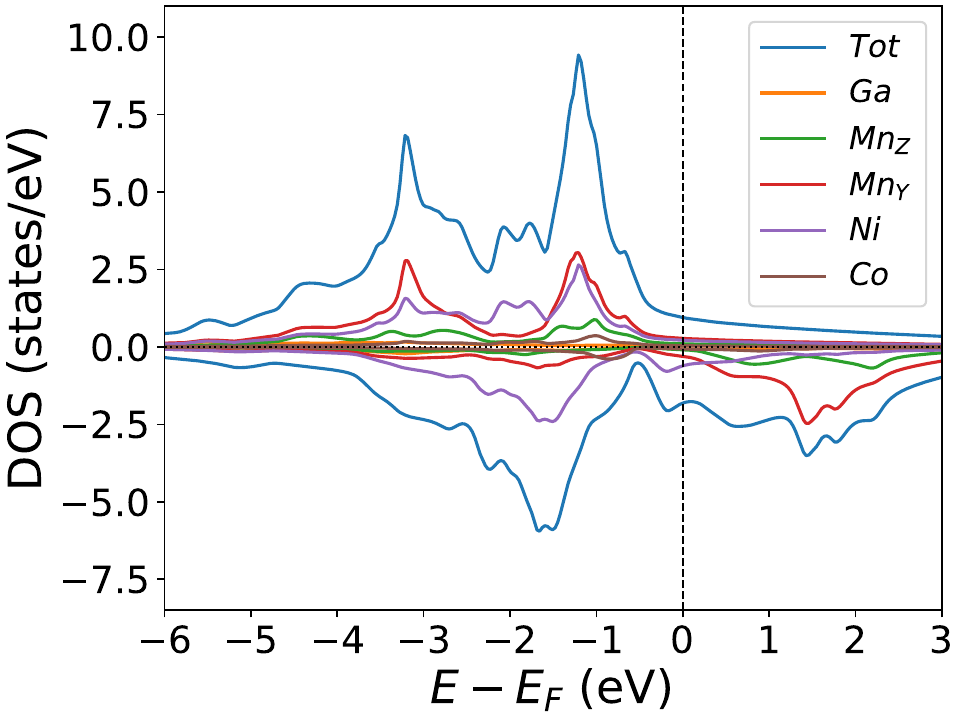}
		\caption{\label{fig:DOS_12_35}DOS of \nmg{0.12,0.35}.}
	\end{subfigure}
	\begin{subfigure}{.24\columnwidth}
		\centering
		\includegraphics[width=1\linewidth]{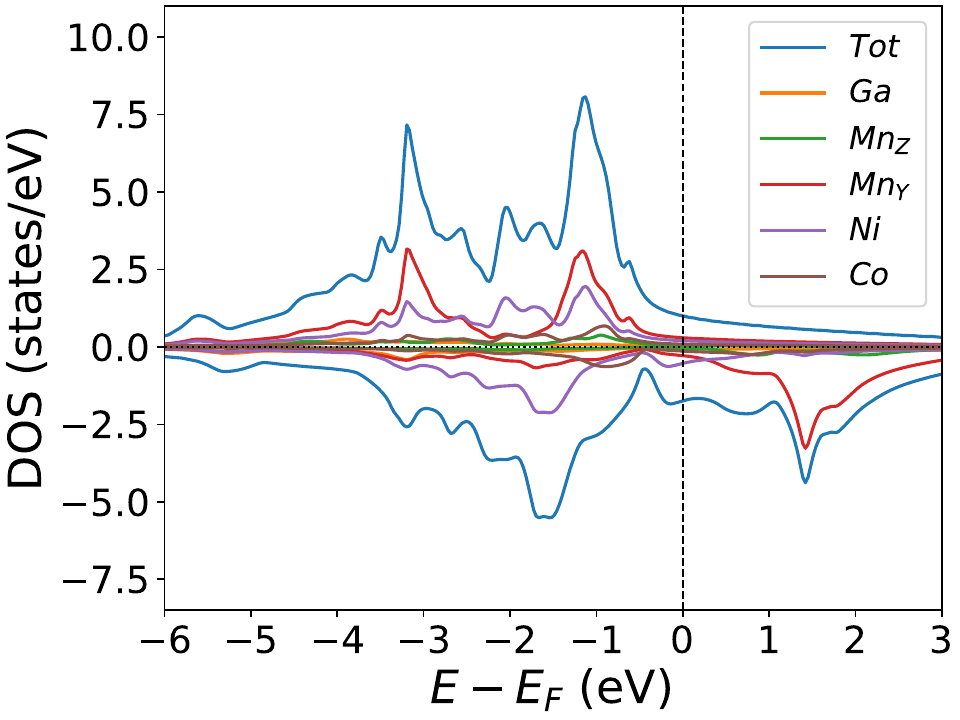}
		\caption{\label{fig:DOS_24_15}DOS of \nmg{0.24,0.15}.}
	\end{subfigure}%
	\begin{subfigure}{.24\columnwidth}
		\centering
		\includegraphics[width=1\linewidth]{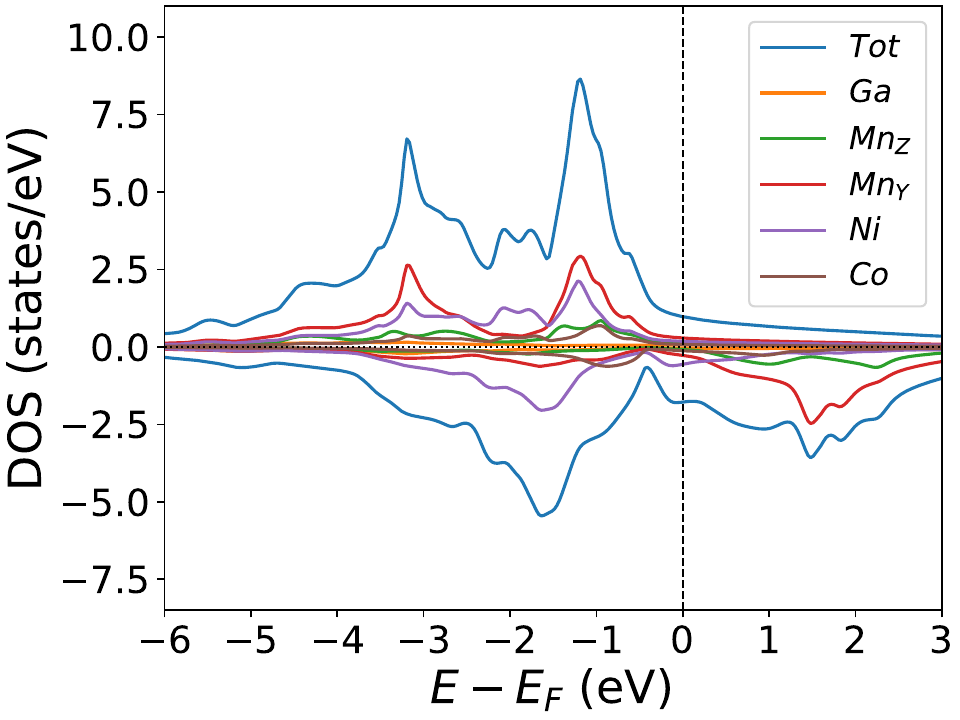}
		\caption{\label{fig:DOS_24_35}DOS of \nmg{0.24,0.35}.}
	\end{subfigure}
	\caption{\label{fig:xydope1}The DOS for completely disordered \nmg{x, z}. 
		Figures ({\ref{fig:DOS_12_15}-\subref{fig:DOS_12_35}}) shows the DOS of $x = 0.12$ and Figures
		({\ref{fig:DOS_24_15}-\subref{fig:DOS_24_35}}) shows the DOS of $x = 0.24$ respectively with ($z$ = 0.15, and 0.35).
		Leftover, $z$ = 0.26 and 0.5 have not been plotted for brevity.}
\end{figure}

\begin{figure}[ht]
	\begin{subfigure}{.24\columnwidth}
		\centering
		\includegraphics[width=1\linewidth]{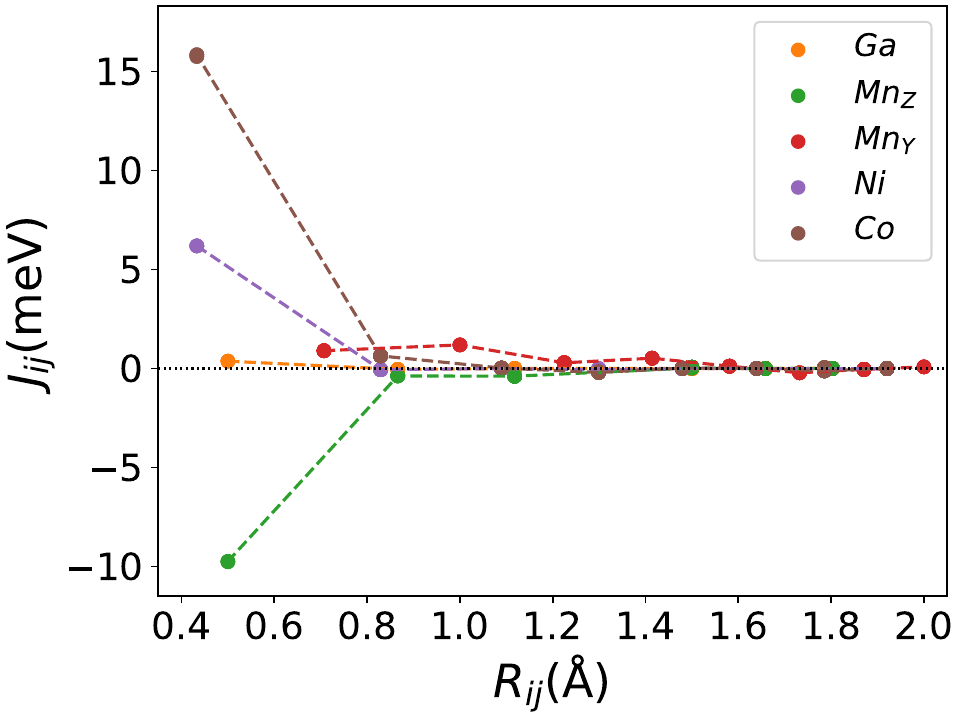}
		\caption{\label{fig:Jij_12_15}$\mathcal{J}_{ij}$ of \nmg{0.12,0.15}.}
	\end{subfigure}
	\begin{subfigure}{.24\columnwidth}
		\centering
		\includegraphics[width=1\linewidth]{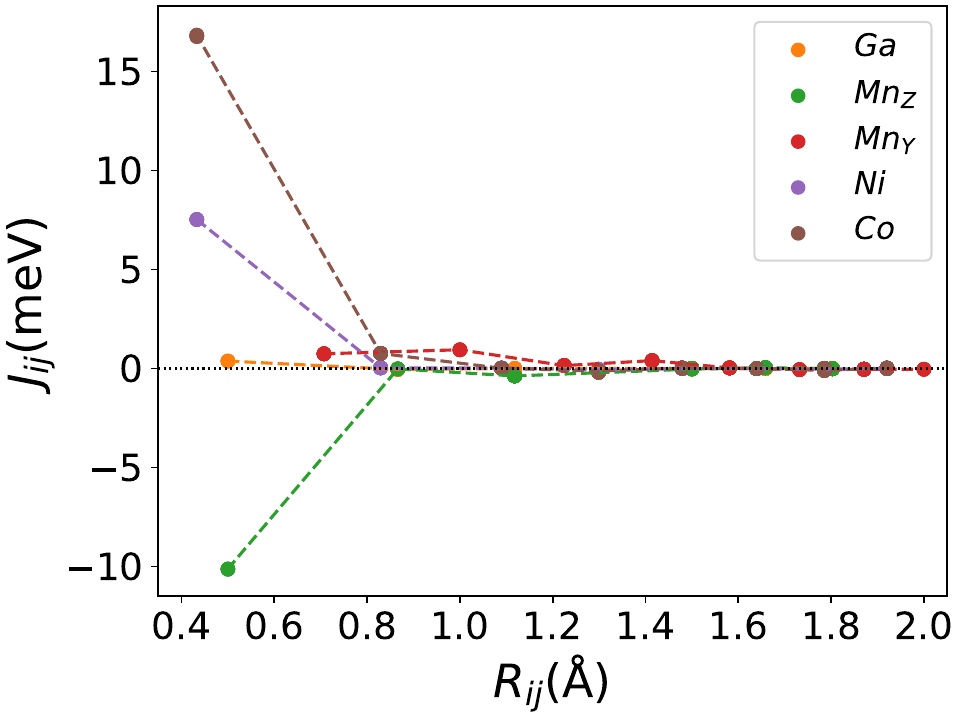}
		\caption{\label{fig:Jij_12_35}$\mathcal{J}_{ij}$ of \nmg{0.12,0.35}.}
	\end{subfigure}
	\begin{subfigure}{.24\columnwidth}
		\centering
		\includegraphics[width=1\linewidth]{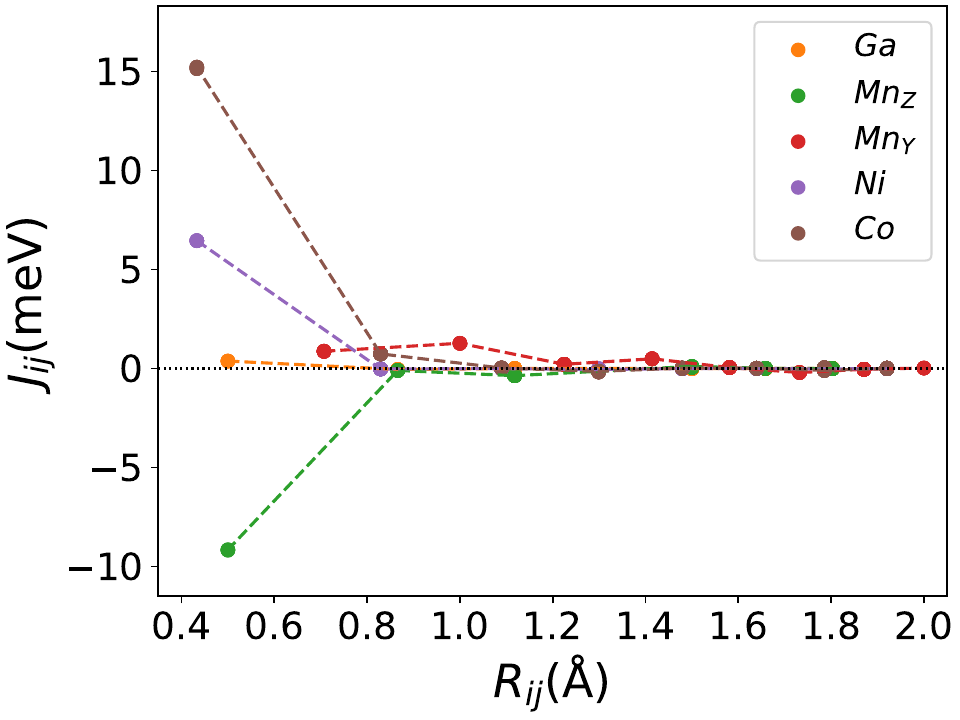}
		\caption{\label{fig:Jij_24_15}$\mathcal{J}_{ij}$ of \nmg{0.24,0.15}.}
	\end{subfigure}
	\begin{subfigure}{.24\columnwidth}
		\centering
		\includegraphics[width=1\linewidth]{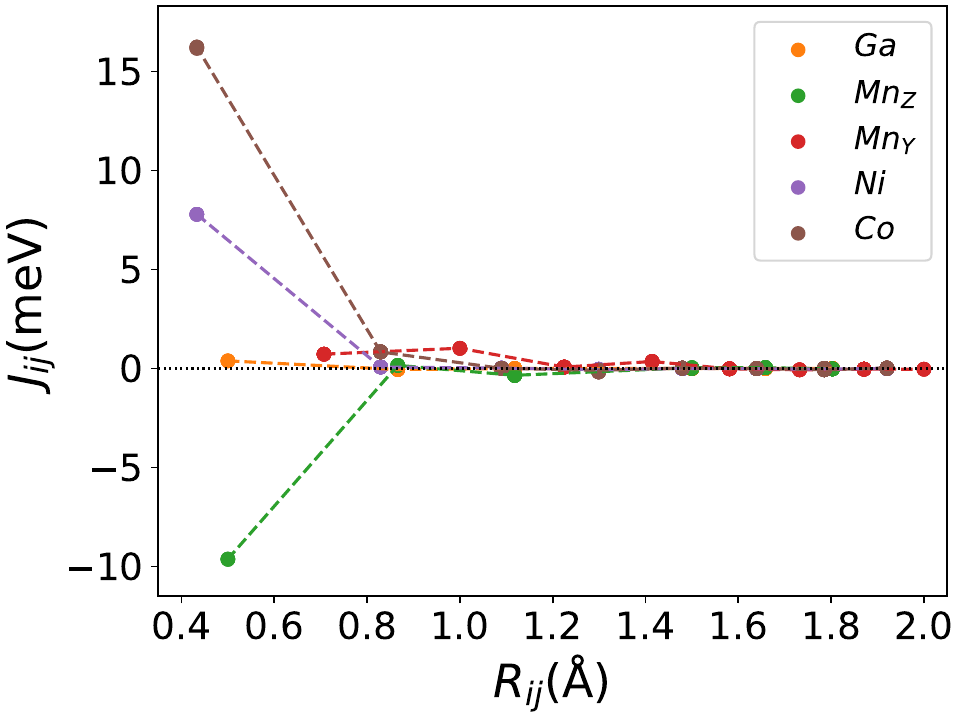}
		\caption{\label{fig:Jij_24_35}$\mathcal{J}_{ij}$ of \nmg{0.24,0.35}.}
	\end{subfigure}
	\caption{\label{fig:xydope2}The Exchange interaction for a completely disordered \nmg{x, z} system.
		Figures ({\ref{fig:Jij_12_15}-\subref{fig:Jij_12_35}}) shows the $\mathcal{J}_{ij}$ of $x = 0.12$ and Figures
		({\ref{fig:Jij_24_15}-\subref{fig:Jij_24_35}}) shows the $\mathcal{J}_{ij}$ of $x = 0.24$ respectively with ($z$ = 0.15, and
		0.35). Leftover, $z$ = 0.26 and 0.5 have not been plotted for brevity.}
\end{figure}

Finally, we study the systems with disorder both at the $X$ and $Z$ sites, i.e., \nicx. Figure (\ref{fig:xydope1}) and Figure
(\ref{fig:xydope2}) represents the electronic and magnetic structures of \nmg{x, z} for  ($x = 0.0, 0.12~\text{and}~0.24$) and ($z = 0.0,
	0.15, 0.26, 0.35~\text{and}~0.5$).

The variation of magnetic moments of \nmg{x, z} are tabulated in Table (\ref{tab:nmg_x-z}).
This doping trend shows the presence of Co at the Ni site($X$ position) and Mn at the Ga site($Z$ position),
which causes the total magnetic moment to increase in accordance with the moments of all other magnetic atoms(Ni, Co and
Mn$_Y$). The moment of a non-magnetic Ga	atom is small and like a constant.
\begin{table*}[ht!]
	\centering
	\begin{tabular}{ccccccccccc}\toprule
		\multicolumn{2}{c}{Concentrations} & {Lattice} & \multicolumn{6}{c}{Magnetic Moment($\mu_B$)} & {$T_C$}  &
		\multirow{2}{*}{$e/a$}
		\\\cmidrule{4-9}
		$x$                                & $z$       & Parameter (au)                               & Ni       & Co
		                                   & Mn$_{Y}$  & Ga                                           & Mn$_{Z}$ & Total & (K)
		\\\toprule

		\multirow{4}{*}{0.12}              & 0.15      & 11.062                                       & 0.37     & 1.08  & 3.51
		                                   & -0.09     & 3.60                                         & 4.90     & 1375  & 7.59 \\
		                                   & 0.26      & 11.066                                       & 0.42     & 1.14  & 3.52
		                                   & -0.09     & 3.60                                         & 5.41     & 1465  & 7.70 \\
		                                   & 0.35      & 11.073                                       & 0.46     & 1.18  & 3.52
		                                   & -0.09     & 3.59                                         & 5.82     & 1538  & 7.79 \\
		                                   & 0.50      & 11.082                                       & 0.52     & 1.26  & 3.54
		                                   & -0.10     & 3.59                                         & 6.52     & 1653  & 7.94 \\\midrule
		\multirow{4}{*}{0.24}              & 0.15      & 11.042                                       & 0.39     & 1.08  & 3.46
		                                   & -0.09     & 3.54                                         & 5.03     & 1389  & 7.53 \\
		                                   & 0.26      & 11.048                                       & 0.44     & 1.13  & 3.47
		                                   & -0.09     & 3.54                                         & 5.54     & 1477  & 7.64 \\
		                                   & 0.35      & 11.054                                       & 0.48     & 1.18  & 3.47
		                                   & -0.10     & 3.54                                         & 5.95     & 1549  & 7.73 \\
		                                   & 0.50      & 11.063                                       & 0.54     & 1.25  & 3.48
		                                   & -0.10     & 3.54                                         & 6.63     & 1662  & 7.88 \\\bottomrule
	\end{tabular}
	\caption{\label{tab:nmg_x-z}Structural, magnetic, and thermodynamic properties of \nmg{x, z} system.}
\end{table*}
Replacing Ga with Mn, however, has an enhanced effect on the total magnetic moment. Especially, Mn$_{Z}=50\%$ gives a
relatively large magnetic moment compared to others.

\begin{figure}[ht]
	\begin{subfigure}{.45\columnwidth}
		\centering
		\includegraphics[width=1\linewidth]{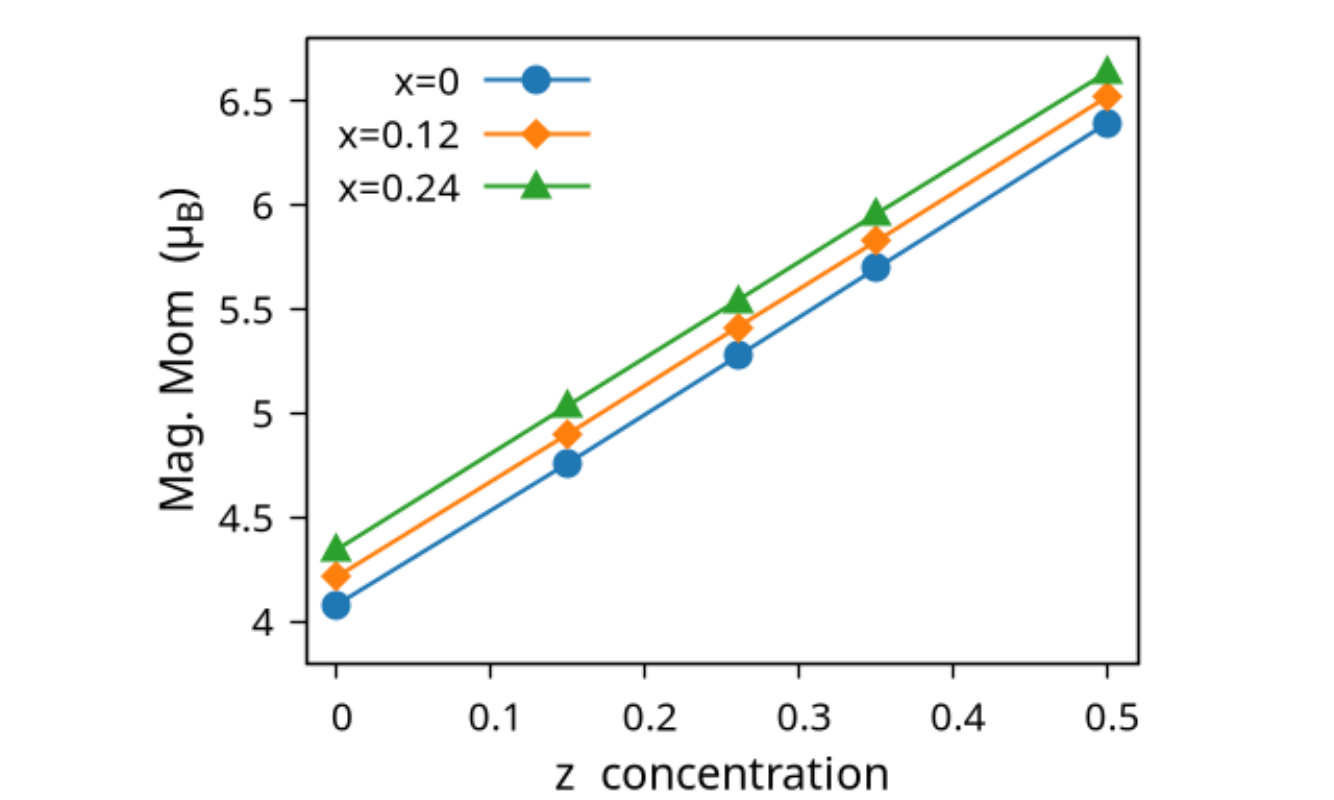}
		\caption{\label{fig:Magnetic_moment_1}Variation of magnetic moment with doping concentration.}
	\end{subfigure}\hfill
	\begin{subfigure}{.45\columnwidth}
		\centering
		\includegraphics[width=1\linewidth]{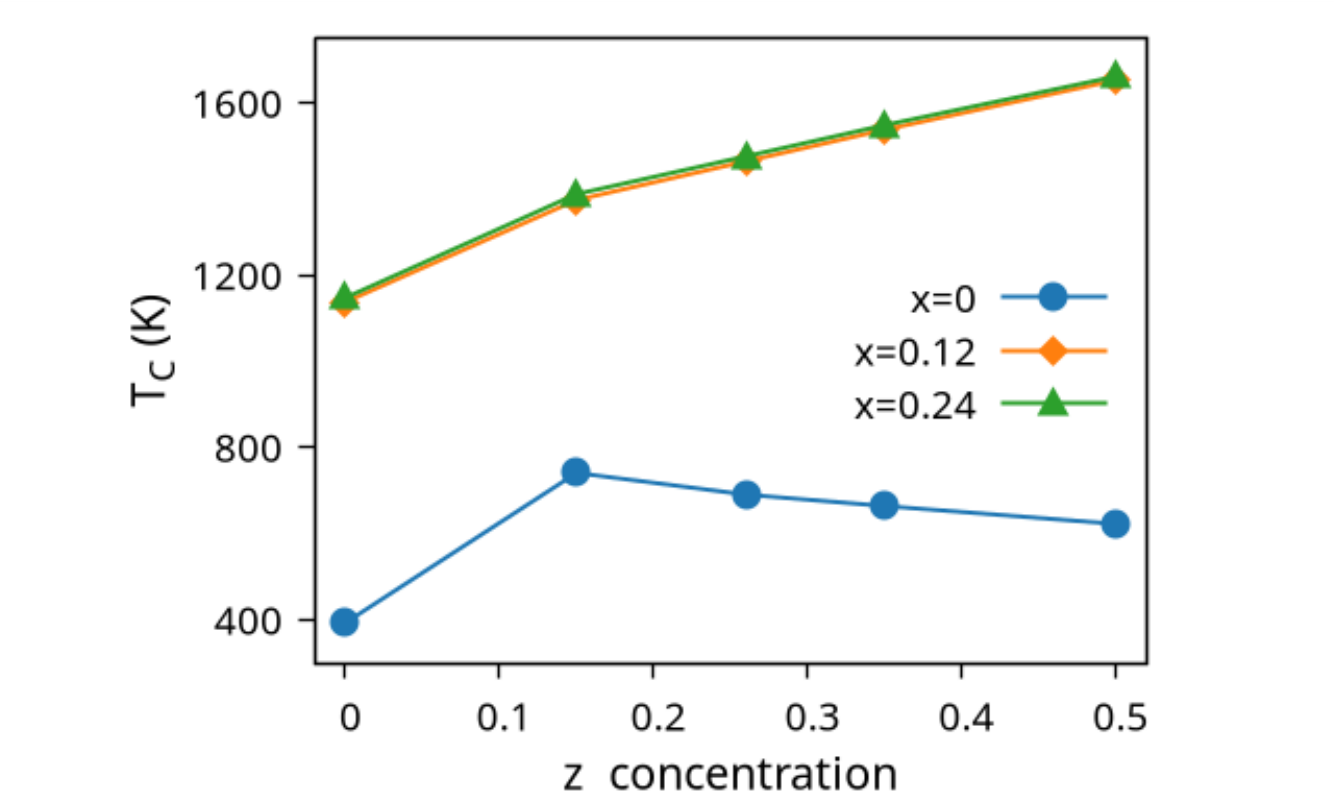}
		\caption{\label{fig:Tc_details}Variation of $T_C$ with doping concentration.}
	\end{subfigure}
	\caption{Variation of  magnetic moment  and $T_C$ of \nmg{x,z}.}
\end{figure}

The variation of the magnetic moment of \nmg{x, z} as a function of doping concentration is shown in Figure
(\ref{fig:Magnetic_moment_1}). The change in magnetic moment of the system is linear within the doping range.
This well-behaved nature is good for tuning and applicability of this system.

In Figure (\ref{fig:xydope2}), we have shown the exchange interactions of \nmg{x, z}.
From the figures, the ferromagnetic interactions are more dominant in the system.
The competing antiferromagnetic interaction(Mn$_Z$-Mn$_Y$) is less than the ferromagnetic interaction(Co-Mn$_Y$).
Notably, the Co-Mn$_Y$ has more ferromagnetic interaction with a value above $\approx$~15 meV.
After initial doping of the $X$ site, the exchange interaction energies are approximately the same for $x$ = 0.12 and 0.24, with
corresponding $z$ concentrations. This could be the cause of the relatively comparable $T_C$ in \nmg{x, z}.

Finally, we have done mean field-based calculations to find the $T_C$ of the systems. Our result shows that doping both $X$ and
$Z$ sites increases the $T_C$. Nevertheless, for \nmg{0, z}, $T_C$ is highest around $z=0.15$, then it starts decreasing. On the other
hand, doping a little in the $X$ site increases the $T_C$ steadily. Our $T_C$ for \nmg{0,0} commensurate with
\cite{Orlandi_2020, Sokolovskiy_2014}. Unfortunately, we are unable to find any literature on the variation of $T_C$
with doping. The variation of $T_C$
is shown in Figure (\ref{fig:Tc_details}) and tabulated in Table (\ref{tab:nmg_x-0}, \ref{tab:nmg_0-y}, \ref{tab:nmg_x-z}) for
\nmg{0~\text{and}~x,0}, \nmg{0, z} and \nmg{x, z} respectively.

The $e/a$ ratio and lattice parameter increase even in the presence of Co at the Ni site since Mn is present at the Ga
site. Here, the $e/a$ ratio and cell volume increase, however the $T_C$ also increases since the presence of Co at the Ni site.
In Table (\ref{tab:nmg_x-z}), we have reported the lattice parameter(au), Curie temperature($T_C$), and $e/a$ ratio values.


\section{Discussions}
\label{sec:discussions}
The lattice parameter increases when Mn is doped at the Ga site ascribed to its larger atomic radius i.e.,
$\text{r}_{\text{Mn}}$(1.61 \AA) $>$ $\text{r}_{\text{Ga}}$(1.36 \AA).
In contrast, doping of Co at the Ni site decreases its lattice constants even though the atomic
radius of Co(1.52 \AA) is larger than Ni(1.49 \AA).
This can be associated with the following reasons: Firstly, the exchange interaction between the
Co and Mn is higher than the Ni and Mn, and this can be well understood from Figures ((\ref{fig:Jij_0_0}),
(\ref{fig:Jij_12_0}-\subref{fig:Jij_24_0}). Secondly, the $d$-$d$ hybridization of Co and Mn is relatively stronger than Ni and
Mn \cite{Tan_2010}. Then the lattice parameter starts to decrease as the Co concentration in the Ni site increases, as shown
in Figure (\ref{fig:lvar}).

In the pristine(\nip) DOS, we observed a pseudogap in the minority channel around $\approx$ 1 eV below the
Fermi-level.
This pseudogap originated
by the hybridization between $3d$ orbitals of Ni and $3p$ orbitals of Ga. The gap is terminated by a peak just below the
Fermi-level originating from the hybridization of the same orbitals and drives the system to Jahn-Teller instability. In \nmg{x, 0} and
\nmg{x, z} systems, the doping at the Ni site stabilizes the Jahn-Teller instability as the peak smears out and the pseudogap
becomes narrower with higher $x$ value, as shown in Figure (\ref{fig:nmg_x-0}\subref{fig:nmg_x0-dos}) for \nmg{x, 0} and Figure
(\ref{fig:xydope1}) for \nmg{x, z}. In \nmg{0, z} systems the pseudogap almost remains the same throughout the doping range as shown in
Figure (\ref{fig:ydope}\subref{fig:nmg_0z-dos}).

The magnetic moments of \nmg{x, 0}, \nmg{0, z}, and \nmg{x, z} are shown in Table (\ref{tab:nmg_x-0}), Table (\ref{tab:nmg_0-y}), and Table
(\ref{tab:nmg_x-z}), respectively. From the tables and Figure (\ref{fig:Magnetic_moment_1}), which shows the variation of the  total
magnetic moment as a function of $x$ and $z$ concentrations, we see the general features:
\begin{enumerate*}[label=(\roman*)]
	\item Magnetic moment is increasing linearly with doping.
	\item Magnetic moment increases faster by doping in $Z$ site than doping in $X$ site. This value is due to the overall magnetic
	moments of $X$ site occupants, i.e., Ni and Co have weaker magnetic moment than Mn$_Z$.
	\item Mn$_Z$ has a higher magnetic moment than the Mn$_{Y}$ site.
	The antiferromagnetic
	interaction between the Mn$_Y$ and Mn$_Z$ atoms is stronger since the inter-atomic distance from Mn$_Y$ to Mn$_Z$ are smaller compared to
	Mn of the same sites(Mn$_Y$ to Mn$_Y$, and Mn$_Z$ to Mn$_Z$).
\end{enumerate*}
The variation in $x$ and $z$ affects the atomic moments in various sites and the total magnetic moments.
General observations are:
\begin{enumerate*}[label=(\roman*)]
	\item Atomistic moments of $X$ sites are most susceptible to $x$ and $z$ concentrations, while $Y$ and $Z$ sites moments remain almost
	unchanged.
	\item Both Ni and Co moment increases with $x, z$ concentration but more with $z$ concentrations.
\end{enumerate*}

The Curie temperature($T_C$) has been calculated using mean field approximations(MFA). As expected, MFA overestimates the $T_C$.
For pure \nip, our calculated $T_C\approx$ 392 K is in good agreement with experimental findings 382 K
\cite{Us_Saleheen_2018}(365 K is reported by \cite{Orlandi_2020, Sokolovskiy_2014}).
For \nmg{x, z}, the qualitative variation of $T_C$ increases monotonously, with the quantitative value matching the previous
findings \cite{Kanomata_2009}. $T_C$ for \nmg{0, z} first increases from 392 K to 741 K for $z=0.15$. Higher doping of Mn$_Z$
decreases the $T_C$ monotonously. For Co-doped systems, both \nmg{x, 0} and \nmg{x, z} increase monotonously and give
approximately the same $T_C$ irrespective of Co concentrations.
To get an understanding of the variation of $T_C$, we have calculated the magnetic pair interaction $\mathcal{J}_{ij}$
as shown in
Figures ((\ref{fig:Jij_0_0}),
(\ref{fig:nmg_x-0}\subref{fig:nmg_x0-jij}), (\ref{fig:ydope}\subref{fig:nmg_0z-jij}), and (\ref{fig:xydope2}))
for \nip, \nmg{x, 0}, \nmg{0, z} and \nmg{x, z} respectively with Mn$_Y$ at the center. When the Co
atom is present, i.e., $x\neq0$(as in \nmg{x, 0} and \nmg{x, z} case), Co and Mn$_Y$ is the dominant interaction. The difference
is that for \nmg{x, 0}, the maximum interaction is almost constant (\ref{fig:nmg_x-0}\subref{fig:nmg_x0-jij}), but for \nmg{x,
	z}, the interaction keeps increasing. The same variation is evident from Figure (\ref{fig:Tc_details}) for $x$~=~0.15 and 0.24.
For \nmg{0, z}, there is a ferro-antiferro competition between Ni-Mn$_Y$ and Mn$_Y$-Mn$_Z$. This brings the $T_C$ down after the
initial increase from pure \nip.
\section{\label{sec:conclusion}Conclusion}
We have investigated the electronic and magnetic properties of \nicx, with $0\leqslant x\leqslant 0.24$, and $0\leqslant z \leqslant
	0.5$. The study of replacing both $X$ and $Z$ is substantially sparse compared to any one of them. We have used DFT and MFA
to study the compound effect of dual-doping. Electronic structures, DOS, and magnetic properties including magnetic
exchange interactions and moments are calculated using DFT as implemented in the SPRKKR package. The Curie temperature($T_C$) is calculated
using MFA. Our calculation shows the existence of strong Mn-Mn antiferromagnetic interaction between Mn at Ga and Mn at its
sub-lattice. Its also noticed that changing concentration at Mn$_{Z}$ does not change the magnetic exchange substantially. This
is an interesting result, as opposed to the findings in the case of $\mathrm{Ni_2Mn_{1+\mathit{x}}Sn_{1-\mathit{x}}}$\cite{Sokolovskiy_2012}.

The $T_C$'s are obtained using MFA simulations using the magnetic exchange interaction values obtained from \textit{ab-initio} calculations.
The numerical results for the pure system are close to that reported by experiments.

We point out the calculations are done in cubic($c/a=1$) phases only. Though, in the dual-doping case, we have shown
the variation of magnetic exchange interaction parameters and magnetic Curie temperatures. These findings will help to
design the functional properties like MCE and shape memory alloys by alloying \nip~suitably.

\section{Acknowledgement}
We acknowledge the High Performance Computing Center (HPCC), SRM IST for providing the computational facility to carry out this
research work effectively.
\bibliographystyle{elsarticle-num}
\bibliography{paper}

\end{document}